\documentclass[floats,floatfix,showpacs,amssymb,prd,twocolumn,superscriptaddress,nofootinbib]{revtex4-1}

\usepackage{amssymb,amsmath,verbatim,mathtools,needspace,enumitem,etoolbox,graphicx,physics,microtype,afterpage,xspace,tabularx,lmodern,multirow}
\usepackage{gensymb}
\usepackage[dvipsnames, usenames]{xcolor}
\definecolor{linkcolor}{rgb}{0.0,0.3,0.5}
\usepackage[unicode, colorlinks=true, linkcolor=linkcolor, citecolor=linkcolor, filecolor=linkcolor, urlcolor=linkcolor, linktocpage, breaklinks]{hyperref}
\usepackage[all]{hypcap}
\usepackage[T1]{fontenc}
\usepackage[utf8]{inputenc}
\usepackage[usenames,dvipsnames]{xcolor}
\hypersetup{colorlinks=true,citecolor=romared,linkcolor=romared,urlcolor=romared}

\definecolor{romared}{RGB}{142,0,28}
\newcommand{\nn}{\nonumber}
\newcommand{\pd}{\partial}
\newcommand{\be}{\begin{equation}}
\newcommand{\ee}{\end{equation}}

\def\be{\begin{equation}}
\def\ee{\end{equation}}
\newcommand{\beq}{\begin{eqnarray}}
\newcommand{\eeq}{\end{eqnarray}}
\newcommand{\ldb}{\lambda_{\mathrm{dB}}}
\newcommand{\ldbbar}{\lambdabar_{\mathrm{dB}}}
\newcommand{\adm}{\mathrm{ADM}}

\usepackage{aas_macros}
\usepackage{makecell}
\usepackage{soul}

\begin{document}

\title{
Relativistic drag forces on black holes from scalar dark matter clouds of all sizes}
\author{Dina Traykova}
%\email{dina.traykova@aei.mpg.de}
\affiliation{Max Planck Institute for Gravitational Physics (Albert Einstein Institute), Am M\"uhlenberg 1, Potsdam-Golm, 14476, Germany}
\author{Rodrigo Vicente}
%\email{rvicente@ifae.es}
\affiliation{Institut de Fisica d’Altes Energies (IFAE), The Barcelona Institute of Science and Technology, Campus UAB, 08193 Bellaterra (Barcelona), Spain}
\author{Katy Clough}
%\email{k.clough@qmul.ac.uk}
\affiliation{School of Mathematical Sciences, Queen Mary University of London,
Mile End Road, London, E1 4NS, United Kingdom}
\author{Thomas Helfer}
%\email{thelfer1@jhu.edu}
\affiliation{Department of Physics and Astronomy, Johns Hopkins University,
3400 N. Charles Street, Baltimore, Maryland, 21218, USA}
\author{Emanuele Berti}
%\email{berti@jhu.edu}
\affiliation{Department of Physics and Astronomy, Johns Hopkins University,
3400 N. Charles Street, Baltimore, Maryland, 21218, USA}
\author{Pedro G. Ferreira}
%\email{pedro.ferreira@physics.ox.ac.uk}
\affiliation{Astrophysics, University of Oxford, DWB, Keble Road, Oxford OX1 3RH, UK}
\author{Lam Hui}
%\email{lh399@columbia.edu}
\affiliation{Center for Theoretical Physics, Department of Physics, Columbia University, New York, NY 10027, USA}
%\date{\today}

\begin{abstract}
We use numerical simulations of scalar field dark matter evolving on a moving black hole background to confirm the regime of validity of (semi-)analytic expressions derived from first principles for both dynamical friction and momentum accretion in the relativistic regime. We cover both small and large clouds (relative to the de Broglie wavelength of the scalars), and light and heavy particle masses (relative to the BH size).
In the case of a small dark matter cloud, the effect of accretion is a non-negligible contribution to the total force on the black hole, even for small scalar masses.
We confirm that this momentum accretion transitions between two regimes (wave- and particle-like) and we identify the mass of the scalar at which the transition between regimes occurs. 
\end{abstract}

\maketitle

%%%%%%%%%%%%%%%%%%%%%%%%%%%%%%%%%%%%%%%%%%%%%%%%%%%%%%%%%%%%%%

%%%%%%%%%%%%%%%%%%%%%%%%%%%%%%%%
\section{Introduction} 
%\noindent{\bf \em Introduction.} 
%%%%%%%%%%%%%%%%%%%%%%%%%%%%%%%%
%
The cold dark matter (CDM) paradigm provides the best explanation to date of the missing mass we observe in galaxies and of large-scale cosmological observations~\cite{Peebles:1982ff,Blumenthal:1984bp,Markevitch:2003at,Bertone:2004pz,Planck:2015mrs,Thomas:2016iav}.
However, the fact that weakly interacting massive particles have so far not been detected directly (despite ongoing attempts~\cite{Schumann:2019eaa}) and the apparent tension of CDM with small-scale (galactic) observations (see, e.g.,~\cite{Weinberg:2013aya,Hui:2016ltb}) has sparked some interest into alternative dark matter (DM) models that still fit large-scale observations, but can show very different behavior on smaller scales.
Light bosonic degrees of freedom (like axions) provide a well-motivated extension of the Standard Model~\cite{Peccei:1977hh,Peccei:2006as,Arvanitaki:2009fg,Marsh:2015xka} and are a possible alternative DM candidate~\cite{Preskill:1982cy,Abbott:1982af,Dine:1982ah,Hu:2000ke, Hui:2016ltb} (see~\cite{Ferreira:2020fam, Niemeyer:2019aqm, Hui:2021tkt} for reviews). 
If such light bosons have masses~$m\lesssim 1 \,\mathrm{eV}$, their de Broglie wavelenght~$\ldb$ is larger than the typical DM inter-particle separation distances in galaxies, and they behave effectively as \emph{classical} waves, exhibiting new phenomenology on scales~$\lesssim \ldb$~\cite{Hui:2021tkt}, which can be astrophysical for the lightest candidates (e.g.,~$\ldb\sim 1 \, \mathrm{kpc}$ for~$m\sim10^{-22} \, \mathrm{eV}$).
Some important manifestations of this wave-like behavior are, e.g., the development of stable long-lived configurations around black holes (BHs)~\cite{Hui:2019aqm,Clough:2019jpm,Bamber:2020bpu,Bamber:2022pbs,Cardoso:2022nzc} due to accretion, or the growth of gravitationally bound clouds powered by superradiance, in the case of spinning BHs~
\cite{Zeldovich:1971,Zeldovich:1972,Starobinskil:1974nkd, Detweiler:1980uk, Cardoso:2005vk, Dolan:2007mj, Herdeiro:2014goa, Arvanitaki:2014wva, Arvanitaki:2010sy,Brito:2015oca,Ghosh:2021uqw,Hui:2022sri}.

Gravitational interactions with compact objects are one of the most promising tools for investigating DM properties, since they do not rely on any additional interactions with the Standard Model. 
Extreme mass-ratio inspirals (EMRIs), in particular, provide an optimal system for studying environmental effects on the inspiral gravitational waveform, since they may complete about~$10^4 - 10^5$ orbits before merger, meaning that small dephasing effects are integrated over long timescales.
Since these kinds of systems typically reside in the center of a galaxy, the smaller object is expected to pass through the DM core, where densities are highest~\cite{Annulli:2020ilw,Annulli:2020lyc,Cardoso:2022vpj}. However, even with next-generation detectors, prospects of observing a signal often rely on enhancements in the density above those in the core, e.g., due to superradiance, accretion of DM spikes, or self-interactions in the DM (see, e.g.,~\cite{Boudon:2022dxi,Baumann:2022pkl,Cardoso:2022whc,Destounis:2022obl,Berti:2022wzk,Foucart:2022iwu,Baryakhtar:2022hbu,Kim:2022mdj}).

Several effects are expected to give rise to dephasing in a binary's gravitational-wave signal when additional matter is present. A key one is dynamical friction (DF): a gravitational drag force due to an overdensity (a ``gravitational wake'') that develops behind a massive object as it moves through a medium. 
%which is the process by which an astrophysical object loses momentum as it moves through a cloud of particles or stars. Due to the gravitational interactions between the larger object and the particles, an overdensity (a ``gravitational wake'') forms behind it, resulting in a gravitational drag force. 
First described by Chandrasekhar~\cite{Chandrasekhar:1943ys} for a nonrelativistic Newtonian perturber moving through a cloud of noninteracting particles, it was then extended to different media (such as fluids~\cite{Ruderman:1971,Rephaeli:1980,Ostriker:1998fa}), different geometries (e.g., spherical or slab-like~\cite{Tremaine:1984,Muto:2011qv,Vicente:2019ilr}), and including relativistic corrections~\cite{Syer:1994vr,Petrich:1989,Barausse:2007ph,Correia:2022gcs}. In the context of scalar field DM, the DF was first computed for a nonrelativistic Newtonian perturber~\cite{Hui:2016ltb}, and then extended to include velocity dispersion~\cite{Lancaster:2019mde,BarOr:2019}, self-gravity~\cite{Annulli:2020ilw,Annulli:2020lyc} (see also~\cite{Wang:2021udl,Chowdhury:2021}) or self-interactions~\cite{Hartman:2020fbg,Boudon:2022dxi}, for binary systems~\cite{Buehler:2022tmr, Tomaselli:2023ysb}, and relativistic perturbers~\cite{Traykova:2021dua,Vicente:2022ivh}. It has recently been shown that relativistic effects can play an important role during the final few orbits of an EMRI, producing a detectable effect on the evolution of the binary~\cite{Speeney:2022ryg}.

Another effect responsible for dephasing in the inspiral of BH binaries is the accretion of matter (and its momentum) onto the BHs as they move relatively to the medium, as originally studied by Bondi~\cite{Bondi:1952ni, Bondi:1944jm}. In most cases it results in a smaller effect than DF (see, e.g.,~\cite{Cardoso:2019rou}), but it is nevertheless important to characterize it, especially in the case of small environments: while the strength of DF increases with the medium (or, more precisely, the wake) size, the effect of momentum accretion is roughly independent of it, and so becomes more important for smaller environments (wakes). In the context of light scalars this accretion was studied by~\citet{Unruh} (see also~\cite{Baumann:2022pkl, Vicente:2022ivh,Boudon:2022dxi, deCesare:2022aoe,deCesare:2023rmg}).

In a previous work~\cite{Traykova:2021dua} some of us characterized the DF effects for \emph{large} scalar DM clouds, by evolving numerically the scalar field around a BH in uniform linear motion, and (based on fluid-media results~\cite{Petrich:1989,Barausse:2007ph}) suggested a phenomenological model for the relativistic corrections which included a pressure-like and Bondi accretion terms. Soon after, analytic expressions for the drag force on the BH were obtained for a similar setup~\cite{Vicente:2022ivh}, suggesting that our phenomenological terms could be removed for very light scalars (i.e.,~$ \gamma M m c/\hbar\ll 1$, where~$\gamma\coloneqq 1/\sqrt{1-v^2}$ is the Lorentz factor of a BH of mass~$M$ and velocity~$v$) by using the analytic expression for Unruh accretion~\cite{Unruh}, without the need of any free parameter. 

In this work, we evolve numerically the scalar field on a moving BH spacetime to confirm that, indeed, the phenomenological relativistic corrections introduced in our previous work can be accounted for by a better modelling of the accretion process. In particular, we find that Unruh accretion captures very well our numerical results for sufficiently light scalars; this is particularly evident for \emph{small} clouds, which we first simulate in this work, where the accretion of momentum gives a comparable (or dominant) effect to DF. 
We also derive the drag force on the BH in the geometrical optics limit (for a cloud of particles following timelike geodesics), which is independent of the particles' spin and, in particular, applies to CDM. The (semi-)analytic expressions that we provide are shown to describe well our numerical results in the different regimes.

Hereafter, we adopt the mostly positive metric signature and use geometrized units in which~$G=c=1$. The scalar field mass will be parametrized by the inverse (Compton) length-scale~$\mu\coloneqq m/\hbar$, and we will often use the dimensionless quantity~$\alpha_{\rm s} = M\mu$ to present our results. For reference: a value of~$\alpha_{\rm s}\sim10^{-2}$ corresponds to a scalar with~$m \sim 10^{-12}\, \rm{eV}$ for a solar-mass BH ($M\sim M_\odot$), or to~$m \sim 10^{-22}\, \rm{eV}$ for a supermassive BH ($M\sim 10^{10}M_\odot$).

%%%%%%%%%%%%%%%%%%%%%%%%%%%%%%%%
%\vspace{0.25cm}
\section{Theory} 
%\noindent{\bf \em Theory.} 
%%%%%%%%%%%%%%%%%%%%%%%%%%%%%%%%
%
We consider a complex scalar~$\varphi$ minimally coupled to gravity described by the action
\begin{equation}
    S=\frac{1}{2}\int \dd^4x \sqrt{-g}\left( \nabla_\mu \varphi^* \nabla^\mu \varphi- \mu^2 |\varphi|^2 \right)\,,
\end{equation}
which results in the Klein-Gordon equation
\begin{equation}
    \left(\Box_g-\mu^2\right) \varphi=0\,.
\end{equation}
We assume a scalar field dilute enough that its backreaction on the spacetime geometry is negligible at leading order; so, effectively we consider that the scalar field evolves on a (vacuum) Schwarzschild geometry. The scalar field energy-momentum tensor is\,
\begin{equation}
    T_{\mu \nu}=\nabla_{(\mu}\varphi^* \nabla_{\nu)}\varphi-\tfrac{1}{2}g_{\mu \nu}\left[\nabla_\delta \varphi^* \nabla^\delta \varphi +\mu^2|\varphi|^2\right]\,,
\end{equation}
where $A_{(\mu} B_{\nu)} \coloneqq \tfrac{1}{2}\left(A_\mu B_\nu +A_\nu B_\mu\right)$ denotes symmetrization.

%%%%%%%%%%%%%%%%%%%%%%%%%%%%%%%%
\section{Coordinate systems} 
%\vspace{0.25cm}
%\noindent{\bf \em Coordinate systems.} 
%%%%%%%%%%%%%%%%%%%%%%%%%%%%%%%%
%
We start with the Schwarzschild metric in isotropic coordinates~$(\bar{t},\bar{x},\bar{y},\bar{z})$, corresponding to the ``BH frame'', which we then boost by a factor~$\gamma$ in the~$\pd/\pd\bar{x}$ direction; the resulting coordinates~$(t',x',y',z')$ correspond to the ``scalar field frame''. By adding a spatially-constant
shift in the~$x'$-coordinate (i.e.,~$x\coloneqq x'-v t'$), we obtain a time-invariant metric in which the BH remains at a fixed coordinate position -- we call this coordinate system, in which we perform the numerical evolution, the ``simulation coordinates''~$(t,x,y,z)$. 

The~$3+1$ ADM decomposition of the Schwarzschild metric in simulation coordinates is
\begin{equation}
\dd s^2=-\alpha^2\dd t^2+\gamma_{ij}(\dd x^i + \beta^i\dd t)(\dd x^j + \beta^j\dd t)\,,
\end{equation}
where the lapse, shift and nonzero components of the spatial metric are, respectively, 
\begin{equation}
\begin{gathered}
\alpha^2 = \frac{AB}{\gamma^2(B - A v^2)}\,,\qquad \beta_i = \delta_{i}^x A v\,, \\ 
\gamma_{xx} =  \gamma^2(B - Av^2)\,,
\qquad \gamma_{yy} = \gamma_{zz} =  B \,,
\end{gathered}
\end{equation}
where $$A \coloneqq \left(\frac{1-M/2\bar r}{1+M/2\bar r}\right)^2\quad \text{and}\quad B \coloneqq \left(1+ \frac{M}{2 \bar r}\right)^4\,,$$ with $\bar r^2 \coloneqq \gamma^2 x^2 + y^2 + z^2$.

While we perform the numerical computations in simulation coordinates, we will present the final gravitational drag forces in the BH frame, where the analytic expressions are more naturally derived. The rate of change of the ADM momentum in the BH frame can be obtained from the one in simulation coordinates using
\begin{equation}
    \dd P^{\adm}_{\bar \mu} = \frac{\partial x^{\mu}}{\partial x ^{\bar \mu}} \dd P^{\adm}_{\mu}\,,
\end{equation}
and~$\dd \bar{t}=\dd t/\gamma$, which results in
\begin{equation}
    \begin{gathered}
        \frac{\dd}{\dd \bar t}P_{\bar x}^{\adm} = \frac{\dd}{\dd t}P^{\adm}_x + v\gamma^2\frac{\dd}{\dd t} P^{\adm}_t\,, \\
        \frac{\dd}{\dd \bar t}P_{\bar t}^{\adm} = \gamma^2\frac{\dd}{\dd t}P^{\adm}_t\,, \quad \; 
        \frac{\dd}{\dd \bar t}P_{\bar{y},\bar{z}}^{\adm} = \gamma \frac{\dd}{\dd t}P^{\adm}_{y,z}\,.
    \end{gathered}
    \label{eqn:coord_transform}
\end{equation}
In Ref.~\cite{Traykova:2021dua} we assumed~$\dd P^\adm_t\approx 0$ for small $\dd P^\adm_x$, but did not justify this further. Whilst, as suggested in Ref.~\cite{Vicente:2022ivh}, this assumption is not necessarily valid, the treatment in Ref.~\cite{Traykova:2021dua} was self-consistent. We provide a clearer and more thorough justification of the assumptions relating to this accretion term in Appendix~\ref{app:Edot}.

%%%%%%%%%%%%%%%%%%%%%%%%%%%%%%%%
\section{Numerical framework} 
%\vspace{0.25cm}
%\noindent{\bf \em Numerical framework.} 
%%%%%%%%%%%%%%%%%%%%%%%%%%%%%%%%
%
Our numerical setup is substantially the same as in Ref.~\cite{Traykova:2021dua}. For completeness, we briefly recap the main elements of the methods in this section, and refer the reader to~\cite{Traykova:2021dua} for more information. The technical details on the numerical grid setup are given in Appendix~\ref{app:numerics}. 

We evolve the scalar field by solving the system
\begin{align}
\partial_t \varphi &= \alpha \Pi +\beta^i\partial_i \varphi \label{eqn:dtphi} ~ , \\ 
\partial_t \Pi &= \alpha \gamma^{ij}\partial_i\partial_j \varphi +\alpha\left(K\Pi -\gamma^{ij}\Gamma^k_{ij}\partial_k \varphi - m^2 \varphi\right)\nonumber  \\
& + \partial_i \varphi \partial^i \alpha + \beta^i\partial_i \Pi \label{eqn:dtPi} ~ ,
\end{align}
on a fixed Schwarzschild geometry, where~$\Pi$ is the conjugate momentum, as defined by Eq.~\eqref{eqn:dtphi}, and~$K$ is the trace of the extrinsic curvature of the background~$K_{ij}$, which in simulation coordinates is simply given by
\begin{equation}
K_{ij} =\alpha^{-1} D_{(i} \beta_{j)} \,,
\label{eqn:extr_curv}
\end{equation}
since the metric is time-invariant.

We set homogeneous initial conditions for the scalar field across the grid, with~$\Re\Pi(t=0)=0$,~$\Re \varphi(t=0)=\varphi_0$,~$\Im \Pi(t=0)=\mu \varphi_0$, and~$\Im \varphi(t=0)=0$. We use an initial amplitude~$\varphi_0=0.1$, but this is an arbitrary choice, since we neglect the backreaction of the field onto the metric and the system is linear, which implies that the final result can be rescaled to different physical densities (assuming that its backreaction remains negligible).

%%%%%%%%%%%%%%%%%%%%%%%%%%%%%%%%
\section{Gravitational drag} 
%\vspace{0.25cm}
%\noindent{\bf \em Gravitational drag.} 
%%%%%%%%%%%%%%%%%%%%%%%%%%%%%%%%
%
To compute the relativistic drag force acting on the moving BH we use the framework developed in Ref.~\cite{Clough:2021qlv}, which allows us to find the leading order term in the scalar rest mass density using the test field approximation. 
The drag force includes both the effects of DF and momentum accretion, and is defined as
\begin{equation}
    F_i\coloneqq \frac{\dd}{\dd t} P_i^{g}\,,\quad {\rm with} \quad P_i^g \coloneqq \int_{\Sigma_{\rm o}} \dd^3x \sqrt{-g}\, t_i^0[g] \,, \label{eqn:force}
\end{equation}
where~$t_\mu^\nu$ is the Einstein's pseudotensor of the total spacetime metric~$g_{\mu \nu}$, which includes the backreaction from the scalar field (see, e.g., Ref.~\cite{Clough:2021qlv}).
The ``curvature momentum''~$P_i^g$ (and the force~$F_i$) depend on the slicing of the spacetime~$\Sigma_{\rm o}(t)$ -- they are well defined once the observers are specified.

For an asymptotically flat spacetime, the ADM momentum can be decomposed into a curvature part and a scalar field part,
\begin{equation}
    P^\adm_i=P_i^g+P_i^\varphi\,, 
\end{equation}
where, at leading order in~$\epsilon$ (with~$|\varphi|\sim \epsilon$),
\begin{gather}
    P_i^\varphi \approx \int_{\Sigma_{\rm o}} \dd^3x \sqrt{-g}\, T_i^0[\varphi,g^{(0)}] \,, \\
    \frac{\dd}{\dd t} P^\adm_i\approx -\int_{\pd {\Sigma_{\rm o}}} \dd S_j\, \alpha T_i^j[\varphi,g^{(0)}]\,.
\end{gather}
Here~$\dd S_j\coloneqq \dd^2x\sqrt{\sigma} N_j$, where~$\sigma$ is the determinant of the induced metric on~$\pd \Sigma_{\rm o}$ and~$N_j$ its outward unit-normal. Thus, at leading order in~$\epsilon$, the drag force can be obtained directly from an evolution of the scalar field on a fixed background -- sidestepping the actual computation of the backreaction on the metric. That is,
\begin{equation}
    F_i\approx -\int_{\pd {\Sigma_{\rm o}}} \dd S_j\, \alpha T_i^j-\frac{\dd}{\dd t} \int_{\Sigma_{\rm o}} \dd^3x \sqrt{-g}\, T_i^0 \,.
\end{equation}

Finally, as shown in Appendix~\ref{app:Edot}, the last expression implies that the steady-state drag force on the BH frame (i.e., in isotropic coordinates) can be computed from
\begin{equation}
F_{\bar i}\approx -\int_{\partial{\Sigma_{\rm i}}} \dd S_j\, \alpha T^j_i-\int_{{\Sigma_{\rm o}} - {\Sigma_{\rm i}}}\dd^3x \sqrt{-g}\, T^\mu_\nu \,^{(4)}\Gamma^\nu_{\mu i}  \,,
\label{eqn:force-extr}
\end{equation}
where the right-hand side is to be evaluated numerically in simulation coordinates, with~$\Sigma_{\rm i}\subset \Sigma_{\rm o}$ a 3-dimensional surface outside the horizon, which contains the curvature singularity.

%%%%%%%%%%%%%%%%%%%%%%%%%%%%%%%%
\section{Analytic expressions} 
%\vspace{0.25cm}
%\noindent{\bf \em Analytic expressions.} 
%%%%%%%%%%%%%%%%%%%%%%%%%%%%%%%%
%
The DF acting on a point-like Newtonian perturber moving at nonrelativistic velocities through a scalar field cloud was first derived in Ref.~\cite{Hui:2016ltb}. These expressions were extended to the case of a BH moving at relativistic speeds in Ref.~\cite{Vicente:2022ivh}, including also the drag force from accretion of momentum. We summarize here the key results, which we will validate against our simulations. 

There are two important dimensionless parameters in this problem: the ratio of the BH size to the reduced (relativistic) Compton wavelength of the scalars
\begin{equation}
    \bar{\alpha}_{\rm s} \coloneqq \frac{M}{\lambdabar_{\rm C}} = \gamma \alpha_{\rm s}\,,
\end{equation}
and the ratio of the characteristic scattering radius to the reduced de Broglie wavelength
\begin{equation}
    \beta \coloneqq \frac{(\frac{1+v^2}{v^2})M}{\ldbbar}= \bar{\alpha}_{\rm s} \left(\frac{1+v^2}{v}\right)\,.
\end{equation}
These parameters control the wave effects, respectively, in the accretion and scattering processes~\cite{Vicente:2022ivh}, with the field behaving as particles in the semi-classical limits~$\bar{\alpha}_{\rm s} \gg 1$ or~$\beta \gg 1$, for large azimuthal numbers~$\ell\gg1$~\cite{Landau:1991wop}.\,\footnote{Note that for~$\bar{\alpha}_{\rm s}\ll1$ only the mode~$\ell=0$ contributes significantly to accretion~\cite{Unruh} and, independently of the value of~$\beta$, the result for the accretion rate does not have a particle analogue. On the other hand, for~$\bar{\alpha}_{\rm s}\gg 1$, the condition~$\beta\gg1$ is necessarily verified and both accretion and scattering are dominated by modes~$\ell \gg 1$, meaning that the scalar field behaves as particles.} Another relevant dimensionless parameter is the ratio 
\begin{equation}
    \Lambda \coloneqq \frac{2 r}{\ldbbar}=2\gamma \mu v r\,,
\end{equation}
which characterizes the radius~$r$ of the cloud (or, more precisely, of the wake) in units of the de Broglie wavelength~$\ldbbar$. The drag force due to accretion of momentum is independent of~$r$ and becomes increasingly important (as compared to DF) for~$\Lambda \lesssim 1$.
Although we can find (semi-)analytic expressions for the steady-state drag force in all regimes, they only take a simple closed form in particular limiting cases; all the expressions in this section are given in the BH frame (isotropic coordinates) and have the form
\begin{equation}
    F_{\bar{x}}\approx -\frac{4 \pi \rho M^2}{v^2} \gamma^2 (1+v^2)^2 \,\big[\mathcal{D}(\bar{\alpha}_{\rm s},\beta,\Lambda)+ \mathcal{A}(\bar{\alpha}_{\rm s},\beta)\big]\,,\label{eqn:force_analytic}
\end{equation}
with~$\rho$ the asymptotic rest-mass density of the medium, and the coefficients~$\mathcal{D}$ and~$\mathcal{A}$ characterizing the contribution from DF and accretion, respectively. 

Let us start with very light scalars~($\bar{\alpha}_{\rm s}\ll1$). 
%\dt{Maybe this is misleading, hence the comments form both referees on this A.1 and B.1. Isn't the first expression valid for any light scalar mass $\alpha_{\rm s}\lesssim 1$, but only large $r$?} In the limit of large scalar clouds~$\Lambda \gg 1$, the drag force on the BH is~\cite{Vicente:2022ivh}
%
\begin{multline}
    F_{\bar{x}}\approx -\frac{4 \pi \rho M^2}{v^2} \gamma^2 (1+v^2)^2 \big\{ \overbrace{\ln \Lambda-1-\Re \Psi(1+i \beta)}^{\mathcal{D}(\bar{\alpha}_{\rm s}\ll 1, \Lambda\gg 1)} \\
    +\underbrace{\tfrac{4 v^3}{(1+v^2)^2}\tfrac{e^{\pi \beta}\pi \beta}{\sinh(\pi \beta)}}_{\mathcal{A}(\bar{\alpha}_{\rm s}\ll 1)\equiv \mathcal{A}_{\rm Unruh}}\big\}\,,\label{eqn:Fd_large-kr}
\end{multline}
where~$\Psi$ is the digamma function. For smaller scalar field clouds, but still much larger than the BH size~($r/M\gg1$), we can also find a simple closed-form expression in the wave limit~$\beta \ll 1$:
\begin{multline}
    F_{\bar{x}}\approx -\frac{4 \pi \rho M^2}{v^2} \gamma^2 (1+v^2)^2 \big\{ \overbrace{{\rm Cin}( \Lambda)+ \tfrac{\sin \Lambda}{\Lambda} - 1}^{\mathcal{D}(\bar{\alpha}_{\rm s}\ll 1, \beta \ll 1)} \\
    +\underbrace{\tfrac{4 v^3}{(1+v^2)^2}\tfrac{e^{\pi \beta}\pi \beta}{\sinh(\pi \beta)}}_{\mathcal{A}_{\rm Unruh}}\big\}\,,\label{eqn:Fd_all-kr}
\end{multline}
where~${\rm Cin}(z) \coloneqq \int_0^z (1 - \cos t) \dd t /t$ is the cosine integral. 
In both expressions the terms in the first line are an extension to relativistic velocities of the DF expression derived in Ref.~\cite{Hui:2016ltb}, while the term in the second line is due to (Unruh) accretion of momentum.\,\footnote{The momentum accretion was actually derived in Ref.~\cite{Vicente:2022ivh}, but it can be obtained directly from the expression for the accretion rate derived by Unruh in~\cite{Unruh}}.

%%%%%%%%%%%%%%%%%%%%%%%%%%%%%%%%%%%%%%%%%%%%%%%%%%%%%%%%%%%%%
\begin{figure}[t!]
	\centering
	\includegraphics[width=0.48\textwidth]{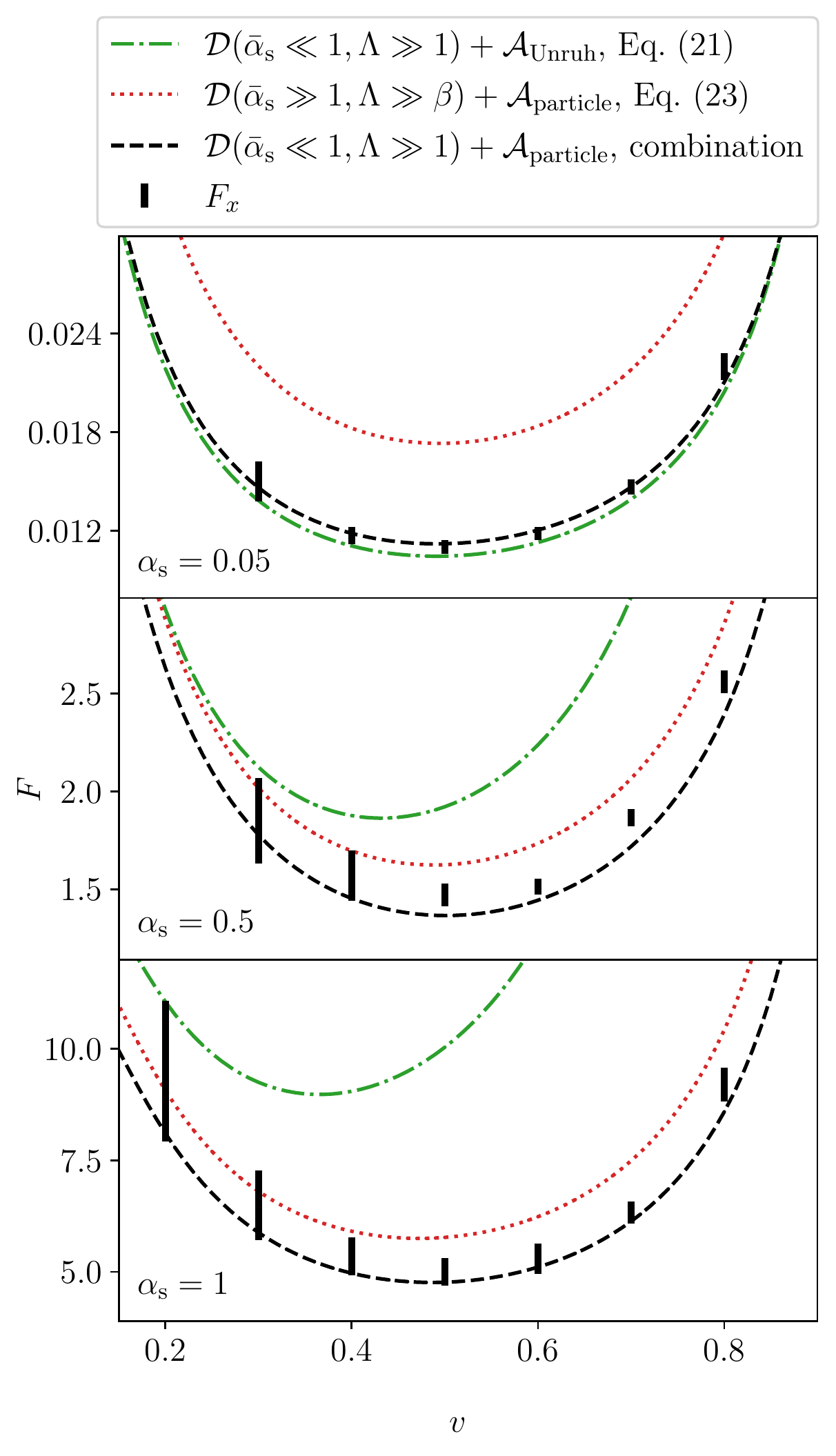}
	\caption{Our numerical results for large clouds~$\Lambda \gg 1$ (shown by 
        the vertical error bars) considering three scalar 
        field masses~$\alpha_{\rm s} = 0.05$,~$0.5$ and~$1$, from top to bottom. The dot-dashed and dotted curves represent, respectively, the analytic expressions for the total force in the light~\eqref{eqn:Fd_large-kr} and heavy~\eqref{eqn:Fd_particle} scalar limits. 
        Unruh accretion reproduces the numerical results in the case of~$\alpha_{\rm s}\ll1$ (top panel), but overestimates them for the other cases where~$\alpha_{\rm s}\lesssim 1$ (middle and bottom panels). Interestingly, the numerical results for these masses seem to be well described by the combination~$\mathcal{D}(\bar{\alpha}_{\rm s}\ll 1,\,\Lambda\gg1)+\mathcal{A}(\bar{\alpha}_{\rm s}\gg 1)$ (shown as black dashed curves), that is, the accretion is in the particle limit, whilst the dynamical friction is in the wave limit.
        }
	\label{fig:accretion}
\end{figure}
%%%%%%%%%%%%%%%%%%%%%%%%%%%%%%%%%%%%%%%%%%%%%%%%%%%%%%%%%%%%%

For heavier scalars~$\bar{\alpha}_{\rm s}\gg 1$, the field behaves as collisionless particles and one recovers geodesic results (cf. Appendix~\ref{app:particle}); using the WKB approximation one can show that, for large clouds~$\Lambda\gg \beta $, the drag force is
\begin{multline}
    F_{\bar{x}}\approx -\frac{4 \pi \rho M^2}{v^2} \gamma^2 (1+v^2)^2 \big\{ \overbrace{\ln \big(\tfrac{\Lambda}{\beta}\big)-1+\chi(v) }^{\mathcal{D}(\bar{\alpha}_{\rm s}\gg 1,\Lambda \gg \beta)}\\
    +\underbrace{\tfrac{v^4}{(1+v^2)^2}\big(\tfrac{b_{\rm cr}(v)}{2 M}\big)^2}_{\mathcal{A}(\bar{\alpha}_{\rm s}\gg 1)\equiv \mathcal{A}_{\rm particle}}\big\}\,,\label{eqn:Fd_particle}
\end{multline}
where the critical impact parameter,
\begin{equation}
    \left(\frac{b_{\rm cr}}{M}\right)^2 \coloneqq \frac{-1+8v^4+\sqrt{1+8 v^2}+4 v^2 (5+2\sqrt{1+8 v^2})}{2 v^4}\, \label{eqn:b_cr}
\end{equation}
separates accretion ($b<b_{\rm cr}$) from deflection ($b>b_{\rm cr}$), and~$\chi(v)$ is a general-relativistic correction to DF given in Appendix~\ref{app:particle}. As expected, the contribution of DF to the total drag force is the same as in Eq.~\eqref{eqn:Fd_large-kr}, in the semi-classical limit~$\beta\gg 1$, modulo general-relativistic corrections contained in~$\chi(v)$, which originate from particles with impact parameters~$b\gtrsim b_{\rm cr}$.
For small clouds we cannot find an analytic expression for the drag force, since the integral~\eqref{eqn:particle_deflect} for the DF contribution must be evaluated numerically.

%%%%%%%%%%%%%%%%%%%%%%%%%%%%%%%%
\section{Numerical results} 
%\vspace{0.25cm}
%\noindent{\bf \em  Numerical results.} 
%%%%%%%%%%%%%%%%%%%%%%%%%%%%%%%%
%
%%%%%%%%%%%%%%%%%%%%%%%%%%%%%%%%%%%%%%%%%%%%%%%%%%%%%%%%%%%%%
\begin{figure}[t!]
	\centering
	\includegraphics[width=0.48\textwidth]{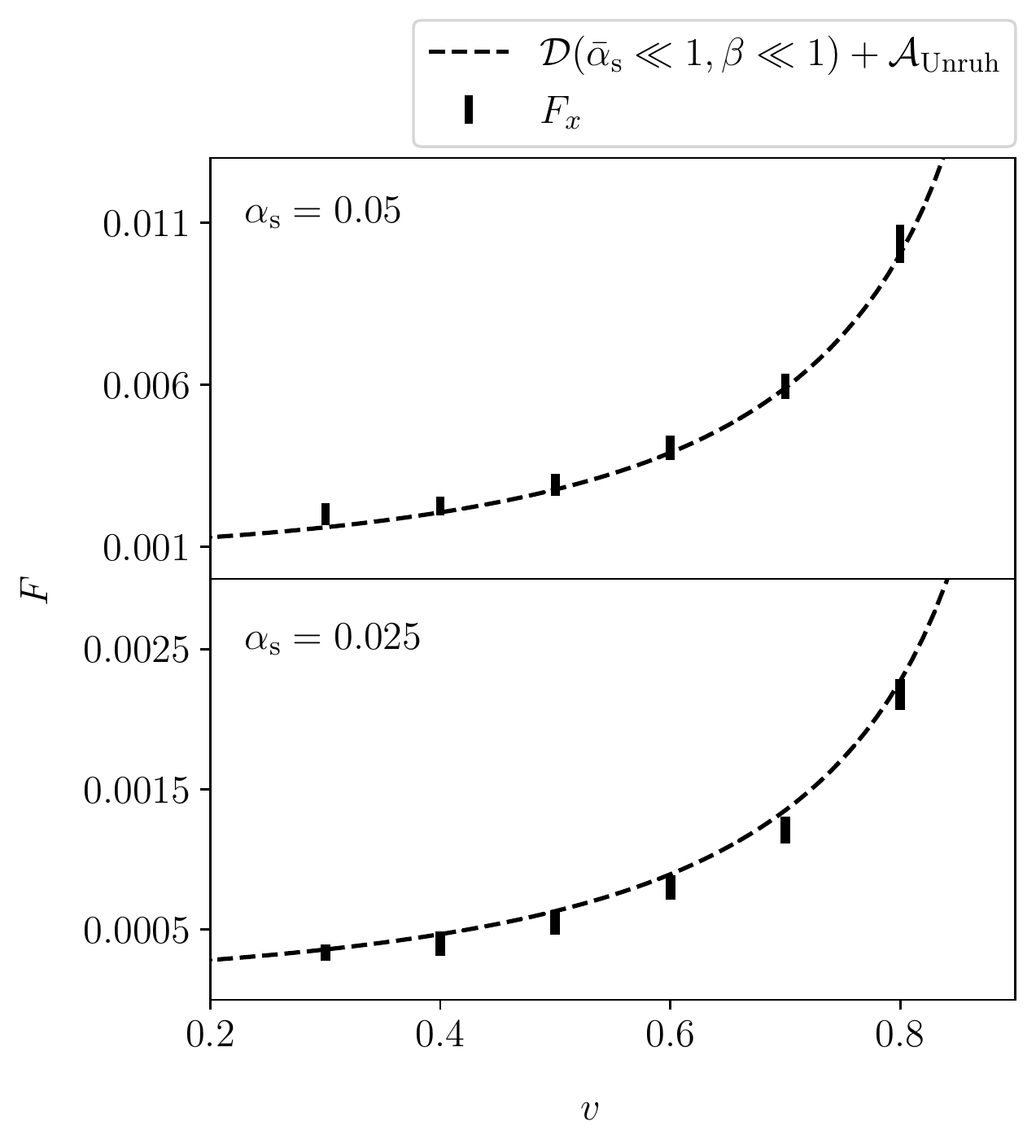}
	\caption{Our numerical results for small clouds~$\Lambda \lesssim 1$ (indicated by the vertical error bars) for the scalar masses~$\alpha_{\rm s}=0.05$ (top panel) and~$\alpha_{\rm s}=0.025$ (bottom panel). The dashed curves show the analytic expression for the total force in Eq.~\eqref{eqn:Fd_all-kr}. We see that the wave-like expressions for the accretion and dynamical friction are an excellent fit.
    }
	\label{fig:small-r}
\end{figure}
%%%%%%%%%%%%%%%%%%%%%%%%%%%%%%%%%%%%%%%%%%%%%%%%%%%%%%%%%%%%%%

We focus our simulations on the boundaries between different regimes of cloud sizes~$\Lambda$ and scalar masses~$\alpha_{\rm s}$, validating the analytic expressions derived in Ref.~\cite{Vicente:2022ivh} for light scalars~($\bar{\alpha}_{\rm s}\ll 1$), and in Appendix~\ref{app:particle} for heavy scalars~($\bar{\alpha}_{\rm s}\gg 1$). These expressions contain no free parameters. Our numerical results for the drag force are obtained extracting all necessary quantities from our simulations to evaluate Eq.~\eqref{eqn:force-extr}. The masses we consider are in the range $\alpha_{\rm s}\in[0.025,1]$ and, so, our simulations could not really probe~$\bar{\alpha}_{\rm s}\gg 1$. The numerical limitation to probe this limit has to do with the increasingly large frequency of the scalar field oscillations, which becomes increasingly harder to resolve.

Figure~\ref{fig:accretion} shows our numerical results along with the plots of the analytic expressions discussed in the previous section for large clouds with different scalar masses~$(\mu r,\alpha_{\rm s})\in \{(45,0.05),(300,0.5),(300,1)\}$. We confirm that for the lightest scalar considered,~$\alpha_{\rm s}=0.05$ (top panel), the total drag force on the BH is correctly accounted for by the expression in Eq.~\eqref{eqn:Fd_large-kr}. We can see from the middle and bottom panels that, at larger masses, the (wave-like) Unruh accretion significantly overestimates the total force.
In these cases, a better approximation for the drag force from accretion is given by~$\mathcal{A}_{\rm particle}$, the second term in Eq.~\eqref{eqn:Fd_particle}, although formally it was derived only in the~$\bar{\alpha}_{\rm s} \gg 1$ limit. Adding the DF from particle DM (general-relativistic) deflection slightly overestimates the force (as can be seen from the red dotted lines in the plots) for the heavier scalars considered~$\alpha_{\rm s}=0.5\,,1$, but we expect it to become more accurate for heavier scalars~($\alpha_{\rm s}\gg 1$).
For intermediate scalar masses~($\alpha_{\rm s}\sim 1$), the best description of the drag force on the BH seems to be provided by a combination of the DF expression derived for~$\bar{\alpha}_{\rm s} \ll 1$ with the (collisionless particle) accretion expression derived for~$\bar{\alpha}_{\rm s} \gg 1$ (black dashed lines). Note that the Unruh accretion is fundamentally wave-like and it only includes the~$\ell=0$ mode, which explains why it becomes an increasingly worse description of the numerical results as we approach~$\bar{\alpha}_{\rm s}\sim 1$; this is mainly due to the no inclusion of higher~$\ell$ modes, which become effective precisely at $\bar{\alpha}_{\rm s} \sim 1$. On the other hand, the DF expression derived for~$\bar{\alpha}_{\rm s} \ll 1$ includes higher~$\ell$ modes and, so, also captures the particle regime. In fact, the only difference between~$\mathcal{D}(\bar{\alpha}_{\rm s}\ll 1, \beta \gg 1)$ and~$\mathcal{D}(\bar{\alpha}_{\rm s}\gg 1)$ are general-relativistic corrections to DM particle deflection, which we speculate to become effective only at~$\bar{\alpha}_{\rm s}\gg 1$.
Our results show that that the transition from wave to particle-like regime occurs for scalar masses~$\alpha_{\rm s}\sim O(0.1)$. However, for the DF this transition appears to be much more gradual than for accretion, and for~$\alpha_{\rm s}\sim 1$ some (general-relativistic) particle effects are still not active.

Figure~\ref{fig:small-r} shows the comparison of our numerical results with the analytic expression~\eqref{eqn:Fd_all-kr} for small clouds~$\Lambda \lesssim 1$ in the wave regime~$\beta \ll 1$.
We confirm that this analytic expression provides an excellent description of our simulations for small clouds~($\mu r=2.5$) with scalar masses~$\alpha_{\rm s} = 0.05$ and~$0.025$. 
For these small clouds the contribution from accretion to the drag force is comparable or more important than DF, and it is then crucial to model it well. While our analytic expressions (based on Unruh accretion) fit well the numerical results for~$\Lambda \sim 1$, the phenomenological ones used in Ref.~\cite{Traykova:2021dua} (based on Bondi accretion) are not good enough (see Appendix~\ref{app:Comparison}).

%%%%%%%%%%%%%%%%%%%%%%%%%%%%%%%%
\section{Discussion} 
\begin{figure}[t!]
	\centering
	\includegraphics[width=0.48\textwidth]{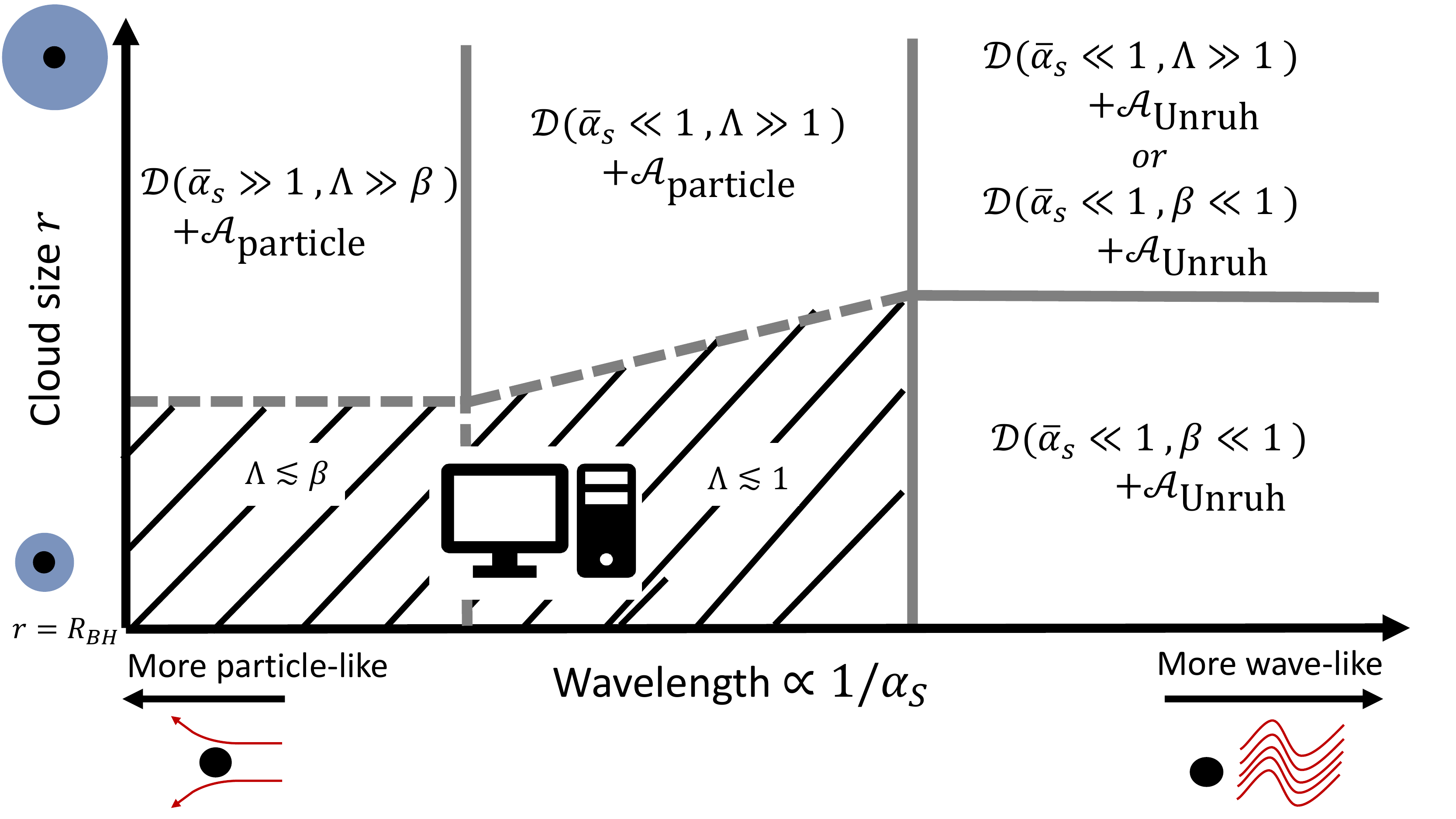}
	\caption{Summary of the different DF and accretion regimes, showing the regions of validity of the expressions in Eqs.~\eqref{eqn:force_analytic}-\eqref{eqn:Fd_particle}.
    In the case of a scalar field wavelength comparable to the size of the BH and smaller and very small scalar field clouds (shaded region on the bottom left of this graph), there is no valid analytic expression for the DF force. Here we can only calculate the drag force on the BH numerically, but as one can see from our results, the analytic expressions give a good order of magnitude estimate even in this regime. 
    } 
	\label{fig:schematic}
\end{figure}

%\vspace{0.25cm}
%\noindent{\bf \em  Discussion.} 
%%%%%%%%%%%%%%%%%%%%%%%%%%%%%%%%
%
%Imaging of supermassive BHs at the centre of M87~\cite{Akiyama:2019cqa} and our own galaxy~\cite{akiyama2022first}, and 
Future detections of EMRIs by space-based detectors like LISA, TianQin and Taiji~\cite{Audley:2017drz, Hu:2017mde, Luo:2015ght, Barausse:2020rsu} will open up new windows on BH environments.
It has been suggested that superradiant clouds may be detectable by LISA~\cite{Macedo:2013qea, Ferreira:2017pth, Hannuksela:2018izj, Zhang:2019eid, Baumann:2021fkf} and, moreover, distinguishable from other environments~\cite{Hannuksela:2018izj,Cole:2022fir}, such as DM spikes, which may cause a similar GW dephasing~\cite{Eda:2013gg, Eda:2014kra,  Yue:2017iwc, Yue:2019ozq, Bertone:2019irm, Hannuksela:2019vip, Edwards:2019tzf, Kavanagh:2020cfn, Coogan:2021uqv, Li:2021pxf}.
To assess the detectability of such effects it is crucial to have a good model of the relativistic drag force acting on BHs in these environments.
In this work we have calculated numerically the force resulting from DF and accretion on a BH boosted through a uniformly dense scalar field medium, and validated the analytic expressions derived from first principles in their different regimes of validity.
In particular, we have shown that the total relativistic drag force has the form given in Eq.~\eqref{eqn:force_analytic}, where the coefficients~$\mathcal{D}$ and~$\mathcal{A}$ contain, respectively, the contributions from DF and momentum accretion. The different regimes for which analytic expression are known are summarized in Fig.~\ref{fig:schematic}.
%%%%%%%%%%%%%%%%%%%%%%%%%%%%%%%%%%%%%%%%%%%%%%%%%%%%%%%%%%%%%
%%%%%%%%%%%%%%%%%%%%%%%%%%%%%%%%%%%%%%%%%%%%%%%%%%%%%%%%%%%%%%

As discussed in Ref.~\cite{Vicente:2022ivh} and confirmed here, the additional ``pressure'' correction that we had considered in our previous work~\cite{Traykova:2021dua} is not necessary when the accretion is correctly accounted for (by using Unruh's expression instead of Bondi's), which becomes particularly evident for the smaller clouds~($\Lambda \lesssim 1$) we have considered here.
Interestingly, we found that at intermediate masses~$\bar{\alpha}_{\rm s}\lesssim 1$ the accretion process transitions between a wave description to a particle one (though the latter expression is formally derived in the~$\bar{\alpha}_{\rm s} \gg 1$ limit).
The general-relativistic effects on particle DF are expected to become effective only at~$\bar{\alpha}_{\rm s}\gg 1$, larger than the masses considered in our simulations.

Having established the methods for extracting and quantifying the relativistic drag forces on BHs moving though scalar field DM clouds, our simulations can now be extended to more complex cases, such as those including DM self-interactions or BH spin, and other fundamental fields such as massive bosons of spin 1 and 2. 
Another interesting extension of our work is to go beyond the test field approximation. All our results were obtained under the assumption that the diluted cloud has a mass much smaller than one of the BH. If that assumption holds, the subleading corrections can be captured by a similar approach to the one used here (but more complicated). Systems for which that assumption does not hold need a different (semi)analytic approach (e.g.,~\cite{Annulli:2020lyc,Cardoso:2023}).

\section*{Acknowledgments}
%\vspace{-0.2in}
%\noindent
We thank V.~Cardoso for helpful conversations. We thank the GRChombo collaboration (www.grchombo.org) for their support and code development work. 
RV was supported by grant no. FJC2021-046551-I funded by MCIN/AEI/10.13039/501100011033 and by the European Union NextGenerationEU/PRTR. RV also acknowledges support by grant no. CERN/FIS-PAR/0023/2019.
KC acknowledges funding from the UKRI Ernest Rutherford Fellowship (grant number ST/V003240/1). 
PF acknowledges support from STFC, the Beecroft Trust and funding from the European Research Council (ERC) under the
European Union’s Horizon 2020 research and innovation programme (Grant Agreement No 693024). 
EB and TH are supported by NSF Grants No. AST-2006538, PHY-2207502, PHY-090003 and PHY-20043, and NASA Grants No. 20-LPS20-0011 and 21-ATP21-0010. 
LH acknowledges support by the DOE DE-SC0011941 and a Simons Fellowship in Theoretical Physics.
The numerical computations presented in this paper used the Sakura cluster at the Max Planck Computing and Data Facility (MPCDF) in Garching, Germany, DiRAC resources under the projects ACSP218 and ACTP238 and resources at the Maryland Advanced Research Computing Center (MARCC). We used the DiRAC at Durham facility managed by the Institute for Computational Cosmology on behalf of the STFC DiRAC HPC Facility (www.dirac.ac.uk), equipment funded by BEIS capital funding via STFC capital grants ST/P002293/1 and ST/R002371/1, Durham University and STFC operations grant ST/R000832/1. DiRAC is part of the National e-Infrastructure. 
The authors also acknowledge the Texas Advanced Computing Center (TACC) at The University of Texas at Austin for providing {HPC, visualization, database, or grid} resources that have contributed to the research results reported within this paper \cite{10.1145/3311790.3396656}. URL: http://www.tacc.utexas.edu. 
For the purpose of Open Access, the author has applied a CC BY public copyright licence to any Author Accepted Manuscript version arising from this submission.

\appendix

\section{Gravitational drag in the BH frame}
\label{app:Edot}

As discussed in the main text, to compare our numerical results with the analytic expressions we need to convert them from simulation coordinates into the BH frame (isotropic coordinates). In a previous work, we assumed that the change in the time-component of the (ADM) four-momentum is negligible compared to the one in the spatial part, and so we ignored the contribution of~$\dd P_t^{\adm}/\dd t$ in Eq.~\eqref{eqn:coord_transform}. While this assumption is not formally valid, we provide here a more careful treatment which confirms that the results presented in Ref.~\cite{Traykova:2021dua} are self-consistent, and thus correct.

Consider three concentric 2-dimensional spheroidal surfaces around the BH: the outer surface~$\partial\Sigma_{\rm o}$, where the ADM mass and overall change in momentum is defined, the inner surface~$\partial\Sigma_{\rm i}$, which is the last well-resolved surface in our simulations before the BH event horizon, and the BH horizon itself~$\partial\Sigma_{\rm BH}$.
The rate of change in the scalar field momentum in the region between~$\partial\Sigma_{\rm BH}$ and $\partial\Sigma_{\rm o}$ is
\begin{multline}
    \frac{\dd}{\dd t}\int_{\Sigma_{\rm o} - \Sigma_{\rm BH}} \dd V \alpha T^0_i  = -\int_{\partial\Sigma_{\rm o}} \dd S_j \alpha  T^j_i  + \int_{\partial\Sigma_{\rm BH}} \dd S_j \alpha  T^j_i  \\
    + \int_{\Sigma_{\rm o} - \Sigma_{\rm BH}} \dd V \alpha T^\mu_\nu \,^{(4)}\Gamma^\nu_{\mu i}\,,
\end{multline}
where~$\dd V\coloneqq \dd^3x \sqrt{-g}/\alpha$.
The flux across the BH horizon in our coordinates is always zero, so we can discard the second term on the right-hand side. 
In the steady state the cloud profile is approximately stationary, so the left-hand term should be small (but not negligible).\footnote{In practice it is large initially, but we want to exclude these changes, which are related to the transient growth in~$\int_{\Sigma_{\rm o} - \Sigma_{\rm BH}} \dd V\alpha T^0_i $ that occurs in our simulations while dynamically evolving towards the steady state. The cloud profile rapidly settles into the steady-state one at smaller~$r$, and more gradually spreads outwards. Whilst there should also be a change in the source term once the outer profile settles into the steady state, we have seen in previous work that this volume integral is dominated by the part at smaller~$r$, and so in practice it also settles quickly to its steady state value~\cite{Clough:2021qlv}.}
In the simulation coordinates, there is always a (small) steady state growth in the momentum of the cloud of~$\dot{M} v$, where~$\dot{M}$ is the steady-state increase in the BH mass over time measured in its rest frame with respect to proper time in that frame. Whilst in that frame this causes an increase in the BH mass and momentum over time, in our coordinates it manifests as the field collecting outside the horizon.
Therefore we have that
\begin{equation}
     \dot{M} v \approx - \int_{\partial\Sigma_{\rm o}} \dd S_j  \alpha T^j_i + \int_{\Sigma_{\rm o} - \Sigma_{\rm BH}} \dd V \alpha T^\mu_\nu \,^{(4)}\Gamma^\nu_{\mu i}  \label{eqn:Pi_ADM1} \,.
\end{equation}
We also need to note that the force acting on the BH is not simply $\dd P_i^{\rm ADM}/\dd t\approx- \int_{\partial\Sigma_{\rm o}} \dd S_j  \alpha T^j_i$, unless $\dd P_i^{\varphi}/\dd t=0$.\footnote{The rate of change of the ADM momentum gives the total force acting on the system (BH + scalar) inside the volume of interest~$\Sigma_{\rm o}$. For a cloud with compact support (and using a sufficiently large~$\partial \Sigma_{\rm o}$), we would have~$\dd P_i^{\rm ADM}/\dd t=0$, meaning that the total force acting on an isolated system vanishes, but the force acting on the BH may still be nonzero if it exchanges momentum with the scalar.}
We need to isolate~$\dd P_i^{g}/\dd t$ (see Eq.~\eqref{eqn:force}), using the fact that in the steady regime
\begin{equation}
    \frac{\dd}{\dd t} P_x^{\varphi} = \frac{\dd}{\dd t} \int_{\Sigma_{\rm o} -\Sigma_{\rm BH}}\dd V \alpha T_x^0  \approx \dot{M} v\,,
\end{equation}
in the simulation coordinates. Thus, in these coordinates the force on the BH is
\begin{equation}
\begin{aligned}
    \frac{\dd}{\dd t} P_x^{g}&= 
    \frac{\dd}{\dd t}P_x^{\rm ADM} - \frac{\dd}{\dd t}P_x^{\varphi} \\
    &=-\int_{\Sigma_{\rm o}-\Sigma_{\rm BH}} \dd V \alpha T^\mu_\nu \,^{(4)}\Gamma^\nu_{\mu x} \,.
\end{aligned}
\end{equation}
For the energy fluxes, the analysis is the same. However, since the background spacetime has a time-like Killing vector, the (volume) source term is exactly zero. Thus, in simulation coordinates, in the steady state we have that
\begin{equation}
    \int_{\partial \Sigma_{\rm o}} \dd S_j \alpha T^j_0  =- \frac{\dd}{\dd t} \int_{\Sigma_{\rm o} - \Sigma_{\rm BH}} \dd V \alpha T^0_0 \approx -\dot{M}/\gamma^2 \,,
\end{equation}
where the~$\dot{M}/\gamma^2$ follows from the fact that~$P^{g}_t = M/\gamma$ in the simulation coordinates, plus the transformation to the BH frame proper time that appears in~$\dot{M}$.
We can see that, using Eq.~\eqref{eqn:coord_transform} to transform the ADM momentum to the BH rest frame, the terms in~$\dot{M}$ cancel exactly such that 
\begin{equation}
    \frac{\dd}{\dd \bar{t}} P_{\bar x}^{\adm}\approx -\int_{\Sigma_{\rm o}-\Sigma_{\rm BH}} \dd V \alpha T^\mu_\nu \,^{(4)}\Gamma^\nu_{\mu x} \,.
\end{equation}
In the BH frame the steady state is such that~$\dd P_{\bar{x}}^{\varphi}/\dd \bar{t}=0$, so~$F_{\bar{x}}\approx \dd P^{\adm}_{\bar x}/\dd \bar{t}$.

In practice in our simulations we cannot resolve the field close to the horizon, but only down to~$\partial\Sigma_{\rm i}$. However, it is easy to check that for a~$\partial \Sigma_{\rm i}$ close enough to the BH event horizon (e.g., Fig. 5 of Ref.~\cite{Traykova:2021dua}), we have
\begin{multline}
    \int_{\partial \Sigma_{\rm i}} \dd S_j \alpha T_i^j+ \int_{\Sigma_{\rm o}- \Sigma_{\rm i}} \dd V \alpha T^\mu_\nu \,^{(4)}\Gamma^\nu_{\mu x}\approx \\
    \approx\int_{\Sigma_{\rm o}-\Sigma_{\rm BH}} \dd V  T^\mu_\nu \,^{(4)}\Gamma^\nu_{\mu x}\,.
\end{multline}
That is, the left-hand side is approximately independent of the choice of~$\Sigma_{\rm i}$ provided it is close to the horizon. This is the quantity that we extract from our simulations and plot as the total BH drag force, in both the current and previous work~\cite{Traykova:2021dua}.

\section{Drag on BHs in particle-like media} \label{app:particle}

For sufficiently heavy scalars~($\bar{\alpha}_{\rm s} \gg 1$), one can use a WKB approximation to recover the geodesic results for the drag force due to the gravitational interaction with a collimated beam of particles~\cite{Vicente:2022ivh}.

Consider the (unbound) time-like geodesics of a Schwarzschild background with energy~$E=\gamma m$ and angular momentum~$L=E v b$, with~$b$ the impact parameter. It is straightforward to find the critical impact parameter~$b_{\rm cr}(v)$, which separates between the particles that are absorbed and the ones that are only deflected by the BH; this is given in Eq.~\eqref{eqn:b_cr}. The drag force due to the accretion of particles is simply given by
\begin{equation}
    F_{\bar{x}}^{\rm accr}= - \rho \gamma^2 v^2 \pi b^2_{\rm cr}(v)\,.
    \label{eqn:particle_accretion}
\end{equation}

The contribution to the DF from the particles that are deflected is found from
\begin{equation}\label{eqn:particle_deflect}
    F_{\bar{x}}^{\rm defl}= - 4\pi \rho \gamma^2 v^2 \int_{b_{\rm cr}}^{b_{\rm max}} \dd b\, b \cos^2 \phi_\infty(b)\,, 
\end{equation}
introducing an infrared cutoff~$b_{\rm max}$ in the impact parameter of the beam, and
where~$\pi-2\phi_\infty$ is the deflection angle, which is given by~\cite{chandrasekhar1998mathematical}
\begin{equation}
    \phi_\infty=\frac{2}{\sqrt{1-6 \tilde{\mu}+2 \tilde{\mu} \tilde{e}}}\left\{ K(k)-F[\tfrac{1}{2} \cos^{-1}(1/\tilde{e}),k]\right\},
\end{equation}
with~$K$ and~$F$, respectively, the complete and the incomplete elliptic integrals of the first kind, with elliptic modulus
\begin{equation}
    k^2 \coloneqq \frac{4 \tilde{\mu} \tilde{e}}{1-6 \tilde{\mu}+2 \tilde{\mu}\tilde{e}}\,.
\end{equation}
The parameter~$\tilde{\mu}\coloneqq M/\tilde{l}$ (with~$\tilde{l}$ the latus rectum) and the eccentricity~$\tilde{e}$ satisfy
\begin{equation}
\begin{aligned}
  \gamma^2 v^2&=\tilde{\mu}\frac{(1-4 \tilde{\mu})(\tilde{e}^2-1)}{1-\tilde{\mu}(3+\tilde{e}^2)}, \\
  \left(\frac{b}{M}\right)^2&=[\tilde{\mu}^2(1-4 \tilde{\mu})(\tilde{e}^2-1)]^{-1}\,.
\end{aligned}
\end{equation}
The general-relativistic effects on the deflection are non-negligible for particles with impact parameter~$b\gg b_{\rm cr}/v$. Thus, for large clouds~$b_{\rm max}/\ldbbar\gg \beta$ ($\gg 1$), one can find a semi-analytic expression
\begin{equation}
    F_{\bar{x}}^{\rm defl}= - \frac{4\pi \rho M^2}{v^2} \gamma^2(1+v^2)^2 \big\{ \ln (\tfrac{b_{\rm max}}{\ldbbar \beta})+ \chi(v) \big\}\,,
\end{equation}
with the (numerically-evaluated) factor
\begin{align*}
    &\chi(v) M^2 \coloneqq \frac{v^4}{(1+v^2)^2} \int_{b_{\rm cr}}^{\frac{\eta(1+v^2)}{4 v} b_{\rm cr}} \dd b\, b \cos^2 \phi_\infty(b) -\ln (\tfrac{\eta v b_{\rm cr}}{4 M} ) \nn\\
    &\qquad \stackrel{\eta =40}{\approx} 3.7+9.24 v^2-12.16 v^3+ 4.79 v^4-\ln (\tfrac{10 v b_{\rm cr}}{M} )\,,
\end{align*}
with an arbitrary~$\eta$ satisfying~$1\ll \eta \lesssim b_{\rm max}/\ldbbar \beta$. The fitting expression in the last line was obtained for~$\eta=40$ and is accurate to~$<1\%$ for all~$v$.
For a cutoff in the radius~$r$ instead (see, e.g., Ref.~\cite{Hui:2016ltb}) the total gravitational drag force on BHs moving through collisionless particle-like media (with no velocity dispersion) is given by Eq.~\eqref{eqn:Fd_particle}.

\section{Comparison to previous results}
\label{app:Comparison}
Here we discuss the differences between the analytic expressions used in this work and the phenomenological ones (based on fluid-like results) used to fit the numerical results in our previous work~\cite{Traykova:2021dua}. 
%%%%%%%%%%%%%%%%%%%%%%%%%%%%%%%%%%%%%%%%%%%%%%%%%%%%%%%%%%%%%
\begin{figure}[t!]
	\centering
	\includegraphics[width=0.48\textwidth]{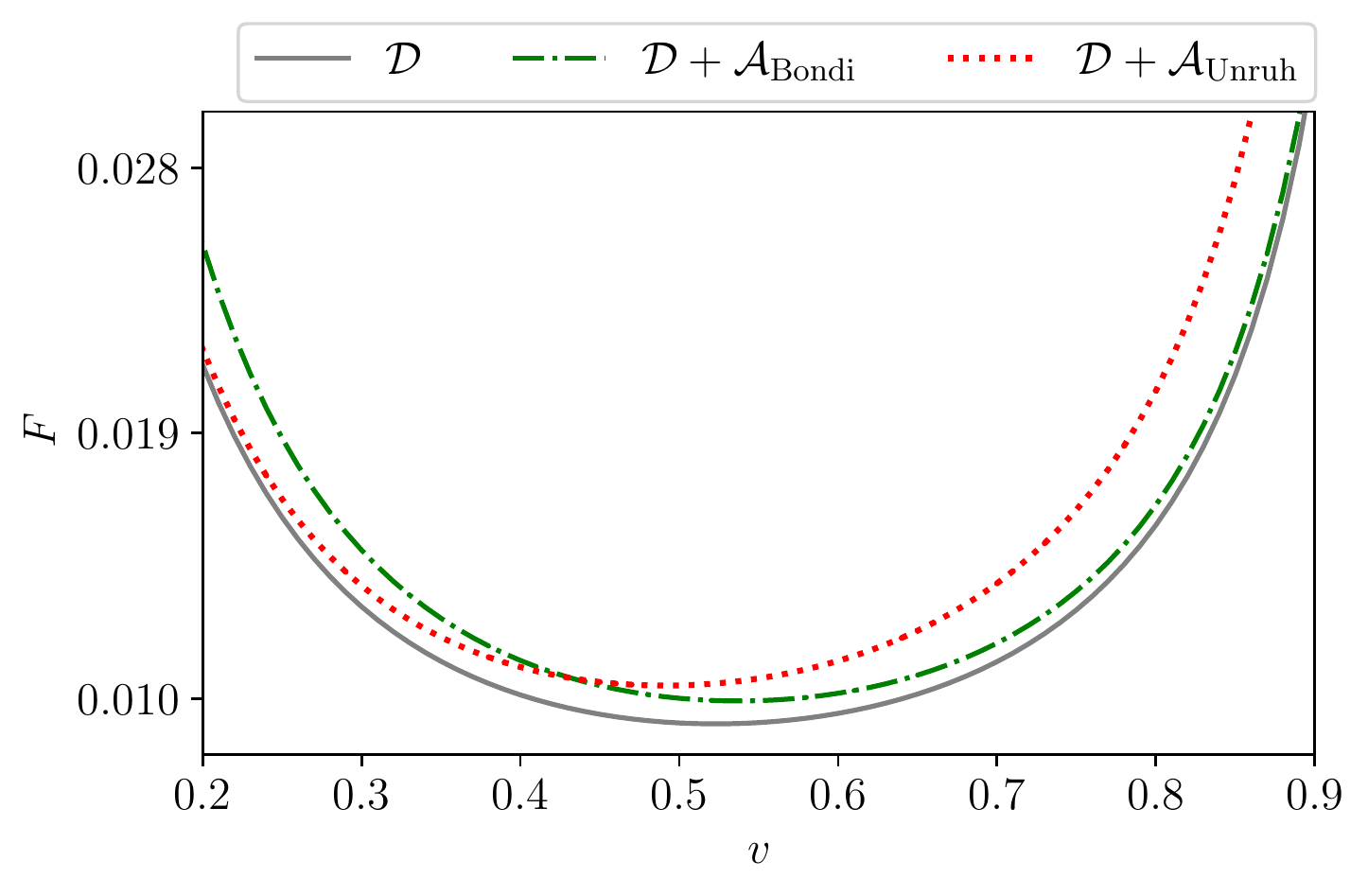}
	\caption{The different contributions to the total drag force on the BH, comparing the addition of Bondi accretion (as used in our previous work) and Unruh accretion (as in \cite{Vicente:2022ivh}) given by the green dashed and red dotted lines, respectively. Comparing with the DF force alone (grey solid line) we can see that Bondi accretion has a non-negligible contribution to the force at low velocities, while Unruh accretion only becomes important at very relativistic speeds.
	All three cases are computed with the same parameters~$\alpha_{\rm s}= 0.05$ and~$\mu r=45$. We set~$\lambda=0.5$ for the~$O(1)$ factor in the expression for Bondi accretion in Eq.~\eqref{eqn:Bondi}.}
	\label{fig:Bondi-vs-Unruh}
\end{figure}
%%%%%%%%%%%%%%%%%%%%%%%%%%%%%%%%%%%%%%%%%%%%%%%%%%%%%%%%%%%%%%
%
There we fit our results using an expression inspired by Bondi accretion~\cite{Bondi:1952ni, Bondi:1944jm}, which describes the accretion of a collisional fluid onto a BH as it moves through the medium and has the form
\begin{equation}
    \mathcal{A}_{\rm Bondi} = \frac{\lambda v^3 }{(1+v^2)^2} (v^2 + c_{\rm s}^2)^{-3/2} \,,
    \label{eqn:Bondi}
\end{equation}
with~$\lambda$ an~$O(1)$ constant factor and~$c_{\rm s}$ the asymptotic sound speed of the medium.
The error bars in our numerical results, due to the long-lived oscillations in the scalar field, were not small enough to reach any definite conclusions on the appropriateness of the description in terms of Bondi accretion. 
In particular, considering~$v\gg c_{\rm s}$, we found that any value of~$\lambda\in [0,1]$ was a good fit to our results.
The increase in the overall (DF + accretion) drag force that we observed compared to the analytic expressions was then attributed to a pressure-like correction to the DF expression~$\mathcal{D}(\bar{\alpha}_{\rm s}\ll1, \Lambda \gg 1)$ in Eq.~\eqref{eqn:Fd_large-kr} (note that we also did not include the Lorentz factor $\gamma$ in~$\Lambda$ and~$\beta$).

Soon after, analytic expressions for the relativistic drag force on BHs moving through light scalars were derived in Ref.~\cite{Vicente:2022ivh}, which found results consistent with ours for the two lightest cases ($\alpha_{\rm s} = 0.05$ and~$0.2$) we had considered.
However, in that reference it was argued that, while Bondi accretion is a good approximation for collisional fluids, the momentum accretion of a light (wave-like) scalar field is better described by an expression derived by Unruh in~Ref.~\cite{Unruh}, which reads
\begin{equation}
    \mathcal{A}_{\rm Unruh} = \frac{4 v^3}{(1+v^2)^2}\frac{e^{\pi \beta}\pi \beta}{\sinh(\pi \beta)}\,.
    \label{eqn:Unruh}
\end{equation}
This adds a non-negligible contribution to the force at relativistic velocities, which removes the need for the (ad-hoc) pressure correction in the DF. It has also the advantage of being fully derived from first principles, containing no free parameters.

As shown in Fig.~\ref{fig:Bondi-vs-Unruh}, the force on the BH from Unruh accretion 
(red dotted line), unlike the one from Bondi's (green dashed line), 
is negligible compared to DF at low velocities, but has a larger effect at high velocities.
%%%%%%%%%%%%%%%%%%%%%%%%%%%%%%%%%%%%%%%%%%%%%%%%%%%%%%%%%%%%%
\begin{figure}[t]
	\centering
	\includegraphics[width=0.48\textwidth]{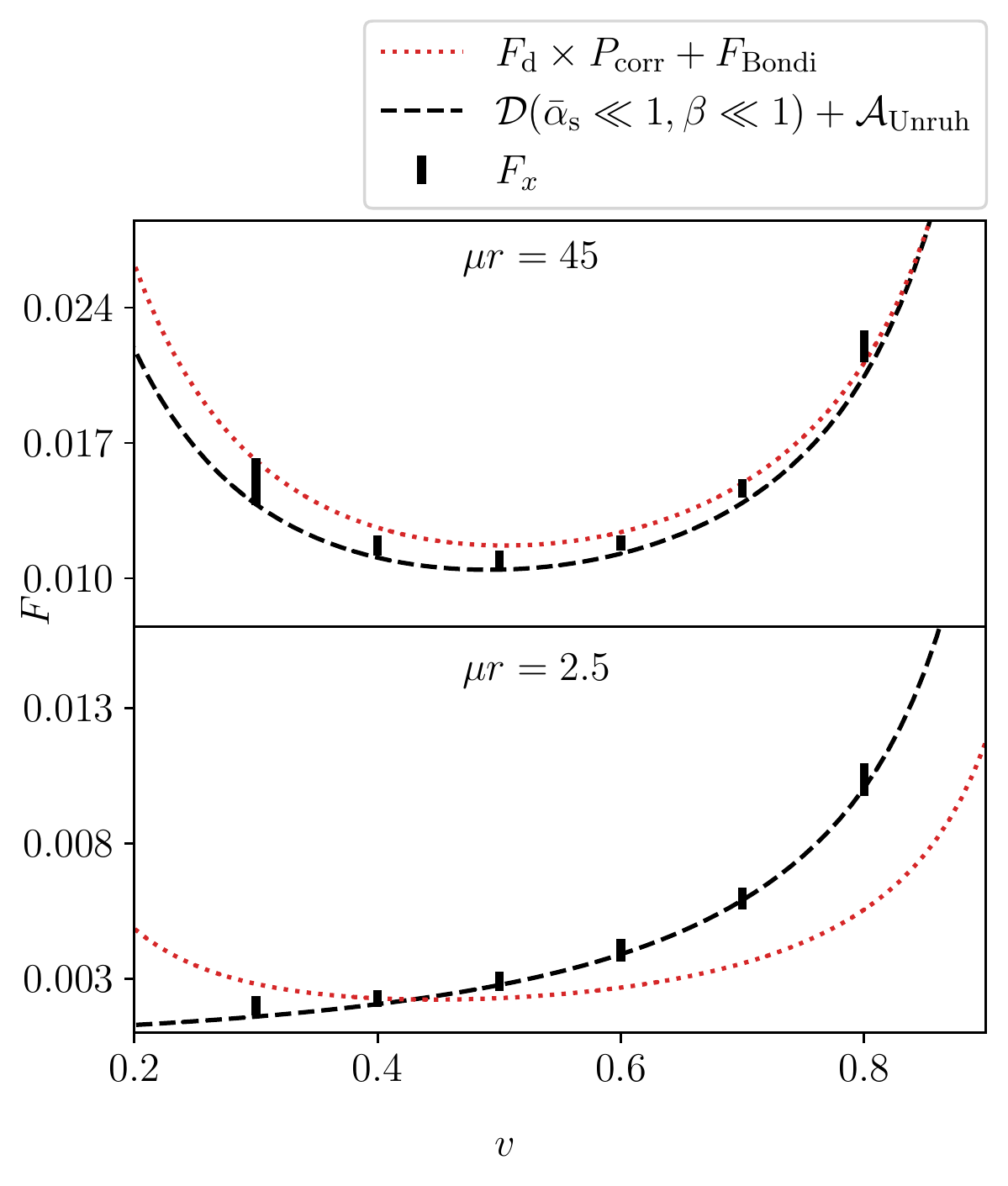}
	\caption{Our numerical results (shown as vertical error bars) for~$\alpha_{\rm s} = 0.05$ in the case of a large cloud ($\mu r=45$) on the top panel, and a small cloud ($\mu r=2.5$) on the bottom panel. The red dotted curve represents the expression in Eq.~(44) of Ref.~\cite{Traykova:2021dua}, including the ``pressure correction'' and using Bondi accretion, and the black dashed curves show the expression for the drag force using Unruh accretion (rather than Bondi) without any phenomenological correction. We see that whilst the previously suggested pressure correction fits the simulation data well at large $r$, it does not provide a good fit for smaller clouds, whereas the expression for Unruh accretion is a good fit in both regimes (for small $\alpha_s$).} 
	\label{fig:frame}
\end{figure}
%%%%%%%%%%%%%%%%%%%%%%%%%%%%%%%%%%%%%%%%%%%%%%%%%%%%%%%%%%%%%%
The Unruh accretion term is particularly relevant for small clouds~($\Lambda \lesssim 1$). In that regime it is also not equivalent to the pressure term that was suggested in Ref.~\cite{Traykova:2021dua}.
This is illustrated in Fig.~\ref{fig:frame}, where the red dotted curves show the expression in Eq.~(44) of Ref.~\cite{Traykova:2021dua}, which includes the ``pressure correction'' factor~$P_{\rm corr} = 1+\kappa p/\rho$ (with~$p$ a scalar field effective pressure and~$\kappa$ another fudge factor) and the Bondi-type accretion. Whilst it provides a good fit for a large cloud~$\Lambda \gg 1$ (top panel), when $\Lambda \lesssim 1$ (bottom panel) the expressions in this paper, which account for the accretion correctly, clearly give a better fit.

\section{Numerical set-up}
\label{app:numerics}

The analytic expression we wish to verify is only valid for small $\beta$, so we take $\alpha_{\rm s} = 0.05$ and $0.025$, although we also compare to previous results for $\alpha_s=0.5$. 
We use the fixed background feature of the open source numerical relativity code $\textsc{GRChombo}$ \cite{GRChombo,Andrade:2021rbd} to solve Eqs.~\eqref{eqn:dtphi} and \eqref{eqn:dtPi} on a fixed metric background in the boosted isotropic coordinates described above.
The scalar field is evolved using the method of lines with its spatial derivatives evaluated using fourth-order finite difference stencils, a fourth-order Runge-Kutta time integration, and a hierarchy of grids with 2:1 resolution.
The value of the metric and its derivatives are calculated locally from the analytic expressions at each point.

Due to the causal structure of the BH in isotropic Schwarzschild coordinates, the metric naturally imposes ingoing boundary conditions at the horizon.
At the outer boundary, we implement nonzero, time oscillating boundary conditions for the scalar field by setting the field to be spatially constant in the radial direction, extrapolating from the values within the numerical domain. This simulates the effects of a roughly constant energy density, but can introduce unphysical effects in very long simulations. These effects can be easily identified by varying the domain size. 
We find that these start to appear on a timescale of roughly twice the size of the numerical domain (in code units), which ultimately limits the time for which the growth of the cloud can be studied.

We choose the size of the simulation domain $L$ such that the approximate boundary conditions that we use do not spoil the solution near the BH in the time necessary to reach a steady state. 
Therefore, we take $L=4096M$, with $9$ $(2:1)$ refinement levels and the coarsest level having $128^3$ grid points. We take advantage of the symmetries of the problem in Cartesian coordinates to reduce the domain to $64^2 \times 128$ points.
This ensures that we maintain a spatial resolution of $\dd x = 0.0625M$ around the horizon of the BH at $r \sim M/2$ and that the BH horizon lies well within the area covered by the finest level.
In order to resolve the temporal oscillations of the field sufficiently on the coarsest level, we use a time step $\dd t_{\rm coarse}$ such that we have at least $32$ time steps per period of oscillation, i.e. $T = 2\pi/\mu > 32\, \dd t_{\rm coarse}$.

The simulations used to extract the numerical results in this work are for the most part (with the addition of~$\alpha_{\rm s}=0.025$) the same as the ones in Ref.~\cite{Traykova:2021dua}. Therefore we do not repeat these here, but instead refer to Appendix A.2 in that paper, where we showed convergence tests in the least and most challenging cases in terms of resolution, ($\mu=0.05$, $v=0.3$) and ($\mu=1$, $v=0.8$), respectively.

\bibliography{DF_small-r}

\end{document}